\documentclass{emulateapj}

\newcommand{\iso}{{\em ISO}}

\newcommand{\mum}{\ifmmode{\rm \mu m}\else{$\mu$m}\fi}

\begin{document}
\slugcomment{ApJSup accepted (Spitzer Special Issue)}

\title{The Infrared Spectrograph\footnote{The IRS was a collaborative 
venture between Cornell University and Ball Aerospace Corporation 
funded by NASA through the Jet Propulsion Laboratory and the Ames 
Research Center.} on the Spitzer Space Telescope}

\author{J.~R.~Houck\altaffilmark{1}, 
T.~L.~Roellig\altaffilmark{2}, J.~van~Cleve\altaffilmark{3}, 
W.~J.~Forrest\altaffilmark{4}, T.~Herter\altaffilmark{1},
C.~R.~Lawrence\altaffilmark{5}, K.~Matthews\altaffilmark{6}, 
H.~J.~Reitsema\altaffilmark{3}, B.~T.~Soifer\altaffilmark{7}, 
D.~M.~Watson\altaffilmark{4}, D.~Weedman\altaffilmark{1},
M.~Huisjen\altaffilmark{3}, J.~Troeltzsch\altaffilmark{3}, 
D.~J.~Barry\altaffilmark{1}, J.~Bernard-Salas\altaffilmark{1}, 
C.~E.~Blacken\altaffilmark{1}, B.~R.~Brandl\altaffilmark{8}, 
V.~Charmandaris\altaffilmark{1,9}, D.~Devost\altaffilmark{1}, 
G.~E.~Gull\altaffilmark{1}, P.~Hall\altaffilmark{1}, 
C.~P.~Henderson\altaffilmark{1}, S.~J.~U.~Higdon\altaffilmark{1}, 
B.~E.~Pirger\altaffilmark{1}, J.~Schoenwald\altaffilmark{1},
G.~C.~Sloan\altaffilmark{1}, K.~I.~Uchida\altaffilmark{1}, 
P.~N.~Appleton\altaffilmark{7}, L.~Armus\altaffilmark{7}, 
M.~J.~Burgdorf\altaffilmark{7}, S.~B.~Fajardo-Acosta\altaffilmark{7}, 
C.~J.~Grillmair\altaffilmark{7}, J.~G.~Ingalls\altaffilmark{7}, 
P.~W.~Morris\altaffilmark{7,10}, H.~I.~Teplitz\altaffilmark{7}
}

\altaffiltext{1}{Cornell University, Astronomy Department, Ithaca, NY
  14853-6801, jrh13@cornell.edu}

\altaffiltext{2}{NASA Ames Research Center, MS 245-6, Moffett Field,
  CA 94035-1000}

\altaffiltext{3}{Ball Aerospace \& Technologies Corp., 1600 Commerce
  St., Boulder, CO 80301}

\altaffiltext{4}{University of Rochester, Department of Physics \&
  Astronomy, Rochester, NY 14627}

\altaffiltext{5}{Caltech, Jet Propulsion Laboratory, Mailcode 169-327,
  Pasadena, CA 91125}

\altaffiltext{6}{Caltech, Palomar Observatory, Pasadena, CA 91125}

\altaffiltext{7}{Caltech, Spitzer Science Center, MS 220-6, Pasadena,
  CA 91125}

\altaffiltext{8}{Leiden University, 2300 RA Leiden, The Netherlands}

\altaffiltext{9}{Chercheur Associ\'e, Observatoire de Paris, F-75014
  Paris, France}

\altaffiltext{10}{NASA Herschel Science Center, IPAC/Caltech, MS
  100-22, Pasadena, CA 91125}

\begin{abstract}
  
  The Infrared Spectrograph (IRS) is one of three science instruments
  on the Spitzer Space Telescope.  The IRS comprises four separate
  spectrograph modules covering the wavelength range from 5.3 to
  38\,\mum\ with spectral resolutions, $R = \lambda/\Delta\lambda \sim
  90$ and $600$, and it was optimized to take full advantage of the
  very low background in the space environment.  The IRS is performing
  at or better than the pre-launch predictions. An autonomous target
  acquisition capability enables the IRS to locate the mid-infrared
  centroid of a source, providing the information so that the
  spacecraft can accurately offset that centroid to a selected slit.
  This feature is particularly useful when taking spectra of sources
  with poorly known coordinates.  An automated data reduction pipeline
  has been developed at the Spitzer Science Center.

\end{abstract} 


\section{Introduction} 

The design of the IRS was driven by the objective of maximizing
sensitivity given the 85\,cm aperture of the Spitzer Space Telescope
\citep{Werner04} and the then-available detectors.  Dividing the
optical trains of the IRS into four separate spectrographs
substantially reduced the complexity and overall cost of the system.
The result is four separate modules, known by their wavelength
coverage and resolution as Short-Low (SL), Short-High (SH), Long-Low
(LL), and Long-High (LH).  The slit widths are set to $\lambda_{\rm
  max}$/85\,cm, where $\lambda_{\rm max}$ is the {\em longest}
wavelength for the module. Further in the geometric limit the
monochromatic slit image covers two pixels. Two Si:As detectors,
128$\times$128 pixels in size, collect the light in the SL and SH
modules, while two Si:Sb detectors of the same number of pixels are
used in the LL and LH modules. In addition to its spectrographs, the
IRS contains two peak-up imaging fields, which are built into the
Short-Low module, and have bandpasses centered at 16\,\mum\ (``blue'')
and 22\,\mum\ (``red'').  A picture of IRS is presented in Figure 1,
and Table 1 gives the parameters of each module.

The two long-slit low-resolution modules were designed for optimum
sensitivity to dust features in the local and distant universe and are
effectively limited in sensitivity by the zodiacal and/or galactic
backgrounds.  The two cross-dispersed high-resolution echelle modules
were designed to achieve the highest possible resolution for the given
array dimensions, and they were optimized for sensitivity to emission
lines.

\begin{deluxetable*}{lcccccc}
\tablecolumns{7}
\tablewidth{0pt}
\tablenum{1}
\tablecaption{Properties of the IRS}
\tablehead{
  \colhead{Module} & \colhead{Array} & \colhead{Pixel Scale (\arcsec)} &
  \colhead{Order} & \colhead{Slit Size (\arcsec)} & 
  \colhead{$\lambda$ (\micron)} & \colhead{$\lambda/\Delta\lambda$}
}
\startdata
Short-Low  & Si:As & 1.8 & SL2 & 3.6$\times$57    & 5.2--7.7\tablenotemark{a}   & 80--128 \\
           &       &     & SL1 & 3.7$\times$57    & 7.4--14.5                   & 64--128 \\
           &       &     &``blue'' peak-up & 56$\times$80 & 13.3--18.7\tablenotemark{b} & $\sim$3 \\
           &       &     &``red'' peak-up &  54$\times$82 & 18.5--26.0\tablenotemark{b} & $\sim$3 \\
Long-Low   & Si:Sb & 5.1 & LL2 & 10.5$\times$168  & 14.0--21.3\tablenotemark{a} & 80--128 \\
           &       &     & LL1 & 10.7$\times$168  & 19.5--38.0                  & 64--128 \\
Short-High & Si:As & 2.3 & 11--20   & 4.7$\times$11.3  & 9.9--19.6                   & $\sim$600 \\
Long-High  & Si:Sb & 4.5 & 11--20   & 11.1$\times$22.3 & 18.7--37.2                  & $\sim$600 \\
\enddata
\tablenotetext{a}{The bonus orders cover 7.3--8.7~\micron\ (SL) and
19.4--21.7~\micron\ (LL).}
\tablenotetext{b}{This is the full width at half maximum of the filter.}
\end{deluxetable*}

In the following sections we describe the instrument, its operation
and calibration, and the reduction of its data using the pipeline
developed at the Spitzer Science Center (SSC).  The Spitzer Observer's
Manual (SOM) provides a more thorough discussion, and it is updated
frequently\footnote{See http://ssc.spitzer.caltech.edu/documents/SOM/
  for the most current version of the SOM.}.

\section{Design, Manufacture, and Ground Testing} 

The IRS contains no moving parts.  The diamond-machined optics are
bolted directly to the precision-machined module housings, and all are
constructed of aluminum so that the assembled modules retain their
focus and alignment from room temperature to their operating
temperature near 1.8 K without any further adjustment.  The focal
plane assemblies are equipped with custom plates to interface with the
module housing and account for the individual dimensions and locations
of the arrays.  Each module contains two flood-illuminating
stimulators to monitor the performance of the arrays.

\begin{figure} 
  \epsscale{1.15}
  \plotone{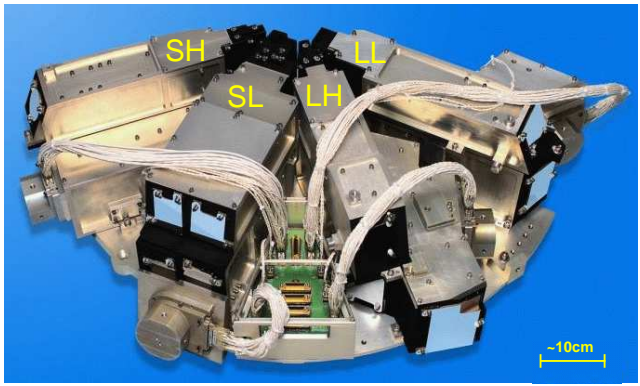}
  \caption{ The Infrared Spectrograph on Spitzer. The four IRS modules, 
    SH, SL, (which includes the peak-up cameras), LH and LL are
    marked. A schematic of the location of the spectrograph slits on
    the Spitzer focal plane is presented in Fig.2 of
    \citet{Werner04}.}
\end{figure}

The low-resolution modules each contain two sub-slits, one for the
first-order spectrum and one for the second-order spectrum.  When a
source is in the sub-slit for the second order, a short piece of the
first order appears on the array; this ``bonus'' order improves the
overlap between the first and second orders.  The high-resolution
modules are cross dispersed so that ten orders (11--20) fall on the
array.

The modules were extensively tested at operating temperature at Ball
Aerospace \citep{Houck00}.  Light from external sources passed through
the Dewar window and a series of neutral density filters at liquid
N$_2$ and He temperatures.  An external optical system projected both
point and extended images from either a blackbody or a monochromator
onto each slit or sub-slit.  These tests verified the resolution,
internal focus and alignment as well as the correct centering of the
spectrum on the arrays.  From the beginning of testing through
integration and launch, the focus and alignment of the modules has
remained constant on a scale of a tenth of a pixel.  The monochromator
tests allowed us to map the wavelength positions roughly on the arrays
in preparation for flight.

Uncertainties in the transmission of the neutral-density filter stacks
prevented a useful determination of the overall sensitivities prior to
launch.  Instead, the sensitivities of the modules were predicted by
an analytic model of the system.  Extensive observations using a
prototype of the Short-High module at the Hale 5m telescope verified
the procedures used in the sensitivity predictions to the 20\% level
\citep{vanCleve98,Smith01}.

When the IRS is operating, all four of its detector arrays are clocked
simultaneously, but it is only possible to capture data from one array
at a time. Two techniques are used in data collection, the double
correlated sampling (DCS mode) and raw data collection (Raw mode, or
``sample up the ramp'').  Science data are collected in Raw mode while
Peak-Up employs DCS. In DCS mode following an initial series of bias
boost and reset frames, each pixel is sampled, and then after a number
of non-destructive spins through the array each pixel is sampled again
via a destructive read and the difference between the two samples is
stored as an 128$\times$128 pixel image. In the RAW mode after the
same initial bias boost, reset frames and first sampling of pixels
there are a number of spin frames followed by a non-destructive read.
This pattern is repeated $n$ times. When all pixels are sampled again
the result is a 128$\times$128$\times n$ cube, where $n$ is the number
of non-destructive reads, and the values are $n$=4, 8, or 16. A final
128$\times$128 image is created by calculating the signal slope of
each pixel of the Raw cube.

\section{In-flight Operation and Calibration} 

Mapping the relative positions of each field of view and the
spacecraft's Pointing Calibration Reference Sensor (PCRS) was the most
critical step of the In-orbit Checkout (IOC) phase.  The positions
must be measured to an accuracy of better than 0$\farcs$14 radial
(0$\farcs$28 for the long-wavelength modules) to meet the 5\%
radiometric requirements.  The positions were measured iteratively,
starting with ground-based estimates and proceeding successively
through ultra-coarse, coarse, and fine focal-plane surveys.  Combined
with the determination of the focus between the telescope and the
slits, this process took place over nearly six of the eight weeks of
the IOC period.  The estimated uncertainties in the final measured
positions of the slits and the peak-up arrays are better than the
requirements in all cases, ranging from 0$\farcs$09 to 0$\farcs$12.

Our knowledge of the internal focus and optical calibration of the
spectral orders were updated in-flight using a combination of
photometric standard stars and emission line objects.  The widths of
the orders were derived from the zodiacal light at maximum intensity,
while order curvatures, tilts, and wavelength solutions were
re-derived from spectral maps of emission line stars such as P Cygni
and planetary nebulae such as NGC 6543, NGC 7027, and SMP 083.  Figure
2 presents the hydrogen recombination spectrum of the Be star $\gamma$
Cas, which provides a good check of the wavelength calibration for
Short-High.

\begin{figure} 
  \epsscale{1.15}
  \plotone{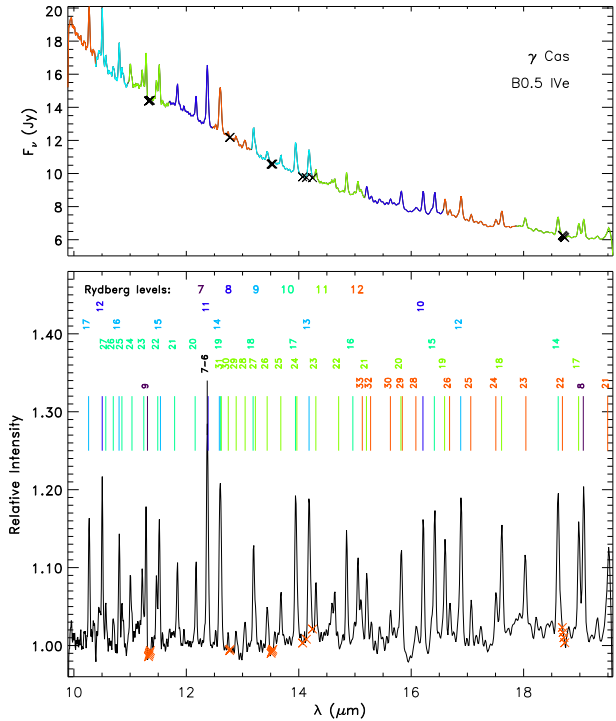}
\caption{A Short-High spectrum of the Be star $\gamma$ Cas.  The 
  lower panel shows a continuum-corrected spectrum with the
  identifications of the hydrogen recombination lines used to check
  the wavelength calibration of this module.  The x's mark the
  location of unusable wavelength elements.}
\end{figure}

Flatfielding and spectrophotometric calibration are determined from
mapping and staring observations of well known standards, including
$\xi$ Dra (HR 6688, K2 III), HR 7310 (G9 III) and HR 6606 (G9 III).
Calibration of the low-resolution modules also used fainter standards
(e.g. HD 42525, A0 V).  \citet{Morris03} provide more details about
the calibration scheme.  \citet{Decin04} discuss the stars and
synthetic spectra used for in-flight spectrophotometric calibration.
We are also verifying the calibration using spectral templates
generated as described by \citet{Cohen03} and observations of
additional standard stars.

The overlaps in wavelength coverage between the various modules and in
orders within each module aid the internal cross-calibration of the
IRS.  These overlaps were also used to search for leaks in the
order-sorting filters by observing ``cold'' and ``hot'' sources in two
overlapping orders.  In the case of the first-order filter for
Long-Low (LL1), which had partially delaminated prior to launch, we
used the combination of Uranus and Neptune as the cold sources and
spectrophotometric standard stars as the hot sources.  At this time
the analysis limits any possible filter leak in LL1 to a maximum of
$\sim$5\% for a Rayleigh-Jeans spectrum.  Further on-going efforts
will refine this limit.

The third largest solar proton flare in the past 25 years began on
2003 October 28, resulting in an integrated proton flux through the
arrays of 1.9$\times$10$^9$ protons cm$^{-2}$, equivalent to the 50
percentile dose expected during the first 2.6 years of the mission.
As a result, the sensitivity of $\sim$4\% of the pixels in Long-High
and $\sim$1\% of the pixels in the other modules have been degraded.
The degree of damage is consistent with the pre-launch damage
experienced by non-flight arrays that were exposed to a beam of 40 MeV
protons at the Harvard Cyclotron Facility.  The IRS is more sensitive
to damage than the other science instruments on board Spitzer because
it is operating at lower background conditions where even a small
increase in dark current has a significant effect.

The measured responsitivity of the system in flight was on average two
times better than the pre-launch model prediction.  The detector noise
measured in the unilluminated parts of the arrays are the same as
measured pre-launch.  Therefore, potentially in the limit of small
signals the sensitivity is on average two times better than the model
predictions.  Most of this increase is due to design margin, an
assumed factor to account of errors and non-modeled effects.  The
assumed margins were a factor of $\sqrt{2}$ for the short wavelength
modules and a factor of 2 for the long wavelength modules.

Updated estimates of sensitivity are available in the current version
of the SOM.  The SSC provides a sensitivity calculator for the IRS
known as SPEC-PET that includes detector noise, source shot noise, and
background shot noise\footnote{See
  http://ssc.spitzer.caltech.edu/tools/specpet/.}.  At the current
time the sensitivity is limited by systematic effects, such as noise
from the flatfield, fringing, and/or extraction edge effects, which
are not included in the calculations of SPEC-PET.  The observed
sensitivities are approximately three times worse than would be
achieved by the above responsivities and detector noise.  It is
anticipated that the realized sensitivities will improve as the system
calibration improves.

\section{Using the IRS} 

\subsection{Spectroscopy} 

The IRS currently has two modes of operation, the spectral ``Staring''
and ``Mapping'' modes. In both cases the target is acquired and then
observed in a fixed sequence starting with SL2 (Short-Low
second-order), SL1, SH, LL2, LL1, LH, skipping any unrequested slits.

The target acquisition method is selected by the observer. One may
simply rely on the blind telescope pointing to point the slits to the
specified position on the sky. A second option is to use the PCRS
\citep{Werner04} to calculate the centroid of a reference star, use
the offset between the star and the science target and move the
science target into the slit. A third and more accurate option is to use
the IRS peak-up cameras to centroid on a 16 or 22\,$\mu$m image of the
science target or a reference star before commencing the requested
spectroscopic observations (see \S 4.2). Using the the PCRS or IRS
peak-up on a reference star requires accurate coordinates for both the
offset star and the science target.

The IRS staring is the more basic mode of operation, and it results in
observing the target at the 1/3 and 2/3 positions along the slit (its
two nod positions), each with the integration time specified for that
slit.  The mapping mode steps the slit parallel and/or perpendicular
to the slit according to the number and the size of the steps
specified by the observer for each slit. The mapping mode does not
perform the 1/3 and 2/3 nod observations that are automatically done
in the staring mode.

Both the staring and mapping modes accept multiple target/position
inputs, as long as all of the positions are within a region two-degree
radius in the sky.  Multiple targets are specified in a ``cluster''
list by either their 1) absolute positions, 2) right ascension and
declination offsets or 3) parallel/perpendicular slit offsets with
respect to an absolute fiducial position. If using IRS in spectral
mapping mode with cluster inputs, parallel/perpendicular slit offsets
can not be used. If the cluster specification is selected, all
sources/offsets in the cluster are observed with the same slit before
proceeding to the next slit. Use of the cluster specification, as
opposed to repeating the same observation one source at a time,
reduces the observatory overhead and can lead to substantial savings
in total ``wall clock'' time.

The SOM describes the IRS observing modes and the various methods of
target acquisition in more detail. In the sections below, we
concentrate on using the IRS peak-up cameras to place a target in the
desired slit, and describe how an observer can use the peak-up fields
for imaging at 16 and 22\,\mum.

\subsection{IRS Peak-up} 

The IRS SL module contains two peak-up imaging fields. Their field of
view is $\sim 55 \arcsec \times 80 \arcsec$, with a scale of 1\farcs 8
per pixel, and their bandpasses are centered at 16\,\mum\ and
22\,\mum\ for the ``blue'' and ``red'' camera respectively (see Table
1).  The IRS peak-up mode enables the placement of a source on a
spectrographic slit or series of slits more accurately than just using
blind pointing of the spacecraft alone.  The telescope blind pointing
has a positional accuracy of $\sim 1 \arcsec$ (1$\sigma$ RMS radial).
An on-board algorithm determines the centroid of the brightest source
in the specified peak-up field and communicates the offsets required
to accurately position the target in the requested slit to the
spacecraft.  As long as the coordinates of a target are accurate
enough to place it on the peak-up imaging field and it is the
brightest object in the field, the IRS will accurately offset to the
selected slits.  The peak-up images are supplied to the observers and
can also be used scientifically if desired.

The allowed ranges of flux densities for blue and red peak-up point
sources are f$_{\rm blue}$=0.8--150\,mJy and f$_{\rm
  red}$=1.4--340\,mJy, respectively.  However, to avoid excessive
integration times and a higher probability of failure, we recommend a
flux of at least 2\,mJy for the blue and 5\,mJy for the red.  The IRS
peak-up algorithm has been optimized for point sources (and indeed
this mode has been the most extensively tested and verified in-orbit),
but an ``Extended Source'' mode is available for sources with
diameters between 5$\arcsec$ and 20$\arcsec$ (e.g. comets).  The
allowed flux density range for extended sources, for either the blue
or red mode, is f$_{\rm ext}$=15--340\,MJy\,sr$^{-1}$.

The observer can specify two peak-up accuracies: ``High'' and
``Medium''.  In the simplest case, when peaking up on the science
target itself, the first move to a slit after peak-up insures
placement to within 0$\farcs$4 (1$\sigma$ RMS radial) of the slit
center for High Accuracy and 1$\farcs$0 for Medium Accuracy.  The High
Accuracy value is driven by the 5\% radiometric accuracy requirement
for the Short-High slit (4$\farcs$7 width).  The IRS section of the
SOM provides details on how these accuracy options apply to multiple
slit positioning after peak-up (e.g., for targets in IRS cluster
mode).

In the event that the science target cannot also serve as the peak-up
target (e.g. if its flux is too low or too high), the ``Offset
Peak-up'' mode gives the option of peaking-up using a nearby source.
The observer is free to use any suitable source within 30$\arcmin$ of
the science target, but the Spitzer planning tool (SPOT) also provides
a list of recommended candidates.  Note that this mode requires
accurate coordinates for both the offset and science target. An IRS
peak-up returns the image of the peak-up target.

The PCRS extends the IRS peak-up capability to optical point sources
with visual magnitudes ranging between $m_{V} = 7$ and $10$.  The only
mode for PCRS currently available provides pointing equivalent to the
High Accuracy mode of the IRS.  The PCRS peak-up does not produce or
return an image.

\subsection{Imaging with the IRS peak-up arrays} 

The peak-up fields in the Short-Low module provide a means of
obtaining images at 16 and 22\,\mum\ (5.4 and 7.5\,\mum\ FWHM
respectively).  The 16\,\mum\ window in particular is interesting
because the other cameras on Spitzer do not cover this wavelength
region, and because this wavelength region was used by the {\it
  Infrared Space Observatory} (\iso) for most of its deep
extragalactic surveys.  Since these surveys probed the properties of
galactic evolution only up to $z\sim 1$, IRS can provide the link
between these past results and the new discoveries which will be made
by Spitzer at $z>2$.

Imaging with the IRS peak-up cameras has not been fully supported for
the first year of operations, but the SSC will provide the ability to
obtain large mosaics in a manner similar to IRAC \citep{Fazio04}
before the end of 2005. However, an interim method called CHEAP
(Cornell High-Efficiency Advanced Peak-up) has been used extensively
to obtain mid-infrared images.  CHEAP is based on an IRS Short-Low
staring observation and uses our knowledge of the offsets between the
center of the peak-up windows and the Short-Low slits in spacecraft
coordinates, along with the standard two-position nodding along the
slit, to place a source target in various positions on the two peak-up
windows.  The user may select any combination of exposure times and
cycles of the Short-Low to obtain an image.  The data are processed by
the standard SSC pipeline as though they are standard Short-Low
spectroscopic observations.  The software that combines and calibrates
the produced images is already available.

As an example, taking a CHEAP image using a standard 60\,sec Short-Low
staring exposure will produce two images at 16 and 22\,\mum\ with an
RMS noise of $\sim 30\,\mu \rm{Jy}$ in only $\sim$720 sec of total
time.  It is interesting to note that the RMS noise of the \iso\ 
observations of the Hubble Deep Field (HDF) was just $\sim 13\,\mu
\rm{Jy}$ at 15\,\mum\ and that the faintest source detected in the HDF
at 15\,\mum\ was 50\% brighter than the sensitivity in the above
example \citep{Aussel99}.  The use of CHEAP to obtain 16 and
22\,$\mu$m images of a sample of high redshift ($z>$1.5) submillimeter
galaxies is presented by \citet{Charmandaris04} and the first
16\,$\mu$m imaging of a Lyman Break Galaxy at $z$=2.79 is discussed by
\citet{Teplitz04}.

\subsection{Planning IRS observations} 

A number of issues should be considered in the design of spectroscopic
observations using IRS. These are addressed in detail in the SOM, but
we briefly mention a few of the more important ones in this section.

Although the infrared background is fainter than the ground based
background by a factor of 10$^{6}$, it may still be necessary to take
an ``off source'' measurement of the background.  The low resolution
modules do this automatically in the sense that both sub-slits are
exposed at the same time so the spectrum from the ``off'' slit can be
subtracted from the ``on'' slit to remove the background.
Alternately, the low-resolution slits are long enough to do the
background subtraction on the slit itself.  However, the
high-resolution slits are too short to subtract the background by
differencing the nod positions.  If the continuum level is important,
off source integrations using the same module are required for the
purpose of background subtraction. SPOT provides an estimate of the
background that is helpful in planning for background subtraction.  As
a general rule, the observers should use the low-resolution modules to
measure continuum and broadband features, and the high-resolution
modules to measure unresolved lines.

The peak-up system works extremely well over its range of parameters.
However, the observer needs to carefully check that the flux of the
peak-up candidate falls within the specified limits of the cameras and
correctly enter it into SPOT.  Furthermore, as already mentioned in
section 4.2 the candidate must be the brightest object within
$\sim$120$''$, to ensure that it will be the one selected by the
peak-up algorithm.  There have been several peak-up failures due to
violations of one or the other of above the requirements.

Observing with Spitzer is unlike most ground based infrared observing.
On the ground the background signal overwhelms the source by many
orders of magnitude.  Therefore, the system noise is set by the shot
noise from the background.  On Spitzer, the noise (N) is dominated by
detector noise for small signals (S), so N$\sim$constant.  However, as
the signal level increases the shot noise in the signal eventually
dominates: N$\propto$S$^{1/2}$.  At the highest signal levels fixed
pattern noise (flat fielding errors, etc.)  are dominant and
consequently N$\propto$S.  At the present time it appears that the
flat fielding term is 1 to 2\%, effectively limiting the maximum S/N
ratio to $\sim$50 to 100.  Similarly at very low signals the flatfield
uncertainty limits our 1$\sigma$ detection to $\sim$0.3\,mJy at
16\,$\mu$m for low backgrounds \citep[see][]{Higdon04a}. The online
S/N estimator includes all but this last term.

\section{Pipeline Data Processing} 

As described in detail in the SOM, IRS observations, either in DCS or
Raw mode, are stored as FITS files of what is called a Data Collection
Event (DCE). A DCE contains all the data obtained by an IRS module
since the most recent destructive read. The IRS Science Pipeline at
the SSC treats each DCE independently. The processing of DCS data,
such as those collected during an IRS peak-up, by the pipeline is
minimal and all processing is performed on-board the spacecraft.
However, for data obtained in Raw mode the pipeline removes basic
instrumental signatures and corrects for variations in spectral
response within and between spectral orders. A number of these steps,
such as corrections/checks for saturation, cosmic rays, dark current
subtraction as well as linearization are performed in the cube level
prior to fitting a slope to the sampled up the detector charge
integration ramp. Others, such as correction for drifts in the dark
current, and stray light or crosstalk between the orders are applied
once the slope image has been created.

The IRS array data supplied in FITS files to the observer are organized
into four categories: Engineering Pipeline Data, Basic Calibrated Data
(BCD), Browse-Quality Data (BQD), and Calibration Data. Additional
files are also available of the masks used in the pipeline processing
as well as processing log and quality assurance files. The Engineering
Pipeline data consist of the raw detector sample images, with some
descriptive information in the FITS file headers. The BCD files are
the fundamental basis for science analysis, with the primary product
being a two-dimensional slope image of each DCE in units of electrons
sec$^{-1}$ pixel$^{-1}$, accompanied by a header containing the
essential programmatic information and the processing, calibration,
and pointing history. Additional BCD data files include uncertainty
images that accompany each slope image and various intermediate images
that do not have all of the pipeline processing corrections applied.
The BCD also include coadded image data for those observations with
more than one ramp at a sky location. It is expected that the BCD
files will be used by the IRS observers as the standard data input for
subsequent publishable science data analysis. As an example, in a
standard IRS staring observation in which the observer has requested 3
cycles of 30sec integrations on a science target with the SH module
will result in 6 DCEs in Raw mode. The Engineering Data pipeline will
process the data, which will have the form of six data cubes (3
integration cycles at each of the two nod positions along the slit),
each 128$\times$128 pixels $\times$ 16 detector reads in size. The BCD
pipeline will produce six 128x128 BCD slope images as well as two
128$\times$128 co-added slope images. The Calibration Data include
measurements of standard calibration objects and can be used with the
BCD to further refine the quality of the science data.

The BQD is designed to provide the observer with an early look at
spectra extracted from 2-dimensional BCD images of point sources. In
general, the BQD spectra are designed to convey the richness of the
observed data and in some cases may be of publishable quality. To
extract spectra from each individual BCD image the post-BCD pipeline
first applies a search algorithm to trace the spectrum in the two
dimensional 128$\times$128 pixel images. Extraction is performed using
an aperture that follows the linear expansion of the point spread
function (PSF) with wavelength. All spectra are extracted as point
sources and no sky background is subtracted before the extraction is
made (see discussion in \S4.1 on how this may affect the extracted
spectrum). The spectra are then photometrically calibrated (in
Janskys) to remove the effects of truncation of the PSF by the slit
aperture and high-order systematic residuals from flatfielding.

Most pipeline operations are now routine, but the need to characterize
the performance of the instrument in-orbit has led to numerous
modifications to the algorithms and adjustments to parameters. Most
important are ongoing efforts to refine the detection and mitigation
of latencies and cosmic ray hits, correct for stray light between the
echelle orders, and remove stray light in Short-Low module originating
from the peak-up apertures. Extraction methods and the final
spectrophotometric calibration of the spectra are also very sensitive
to the pixelation of PSF at each wavelength, and as a result both the
extraction and calibration will require some fine-tuning as a result
of what has been learned from in-flight measurements. A detailed
user's guide to the SSC pipeline and its products is in preparation.

\section{Spectroscopy Modeling, Analysis and Reduction Tool} 

The IRS Spectroscopy Modeling, Analysis and Reduction Tool (SMART) is
an IDL software package used to reduce and analyze IRS data from the
four modules and the two peak-up arrays\footnote{Cornell maintains
  information on their website for SMART and updates it regularly.  See
  http://isc.astro.cornell.edu/smart.}.  SMART is designed to operate
on the basic calibration data (BCD) delivered by the SSC pipeline and
can be run both interactively and in batch mode.  It provides a suite
of quick-look routines that enable observers to assess the quality of
their data.

SMART extracts spectra from either individual or coadded BCDs, and
these can be weighted by a pixel mask which characterizes the noise in
the data.  The extraction routines include column extraction of a
point source scaled to the instrumental PSF, fixed-aperture extraction
for extended sources and a weighted Gaussian extraction for point
sources.

Several options are available to optimize the sky subtraction.  The
spectra from the individual modules can be combined to produce a
single spectrum from 5--38\,\mum.  Spectral analysis, including line,
black-body, template and zodiacal light fitting, photometry,
defringing and dereddening are also available.  The spectral analysis
package is based on the \iso\ Spectral Analysis Package \citep[ISAP,
][]{Sturm98}.  SMART includes packages developed by the Spitzer Legacy
teams. The C2D team provided a number of defringing tools while the
FEPS team has contributed IDP3, an image analysis software package
originally developed for the NICMOS instrument of the Hubble Space
Telescope.

\citet{Higdon04b} provide more details about the SMART package.  It
will become publically available in 2004 September.

\section{Sample Results} 

\begin{figure} 
  \epsscale{1.15}
  \plotone{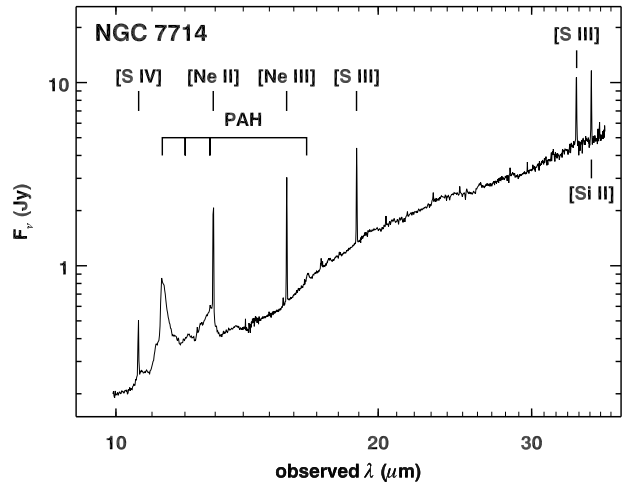}
  \caption{A high-resolution spectrum of the starburst galaxy NGC\,7714 
    (using both Short-High and Long-High) with the detected features
    marked (shifted for z=0.0093).  This spectrum was obtained with 240
    seconds of integration in Short-High and Long-High (each).
    \citet{Brandl04} describe these observations in more detail.}
\end{figure}

The articles in this special Spitzer issue provide many examples of
the quality of the spectra from the IRS.  In addition to the spectrum
of $\gamma$ Cas in Figure 2, we show examples of one low-resolution
spectrum and one high-resolution spectrum, both of galaxies.  All data
presented have been processed and extracted from individual DCEs using
the IRS pipeline at the SSC. Figure 3 presents a spectrum of NGC 7714
using Short-High and Long-High (240 seconds integration each).
\citet{Brandl04} describe this spectrum in more detail, but one can
immediately note the many forbidden lines and dust features present.
Figure 4 presents the spectrum of UGC 5101 obtained using the
low-resolution modules.  The integration times were 12 seconds in each
Short-Low sub-slit and 28 seconds in each Long-Low sub-slit, and this
was sufficient to show a rich spectrum with gas, dust, and ice
features.  See \citet{Armus04} for more information on this spectrum.

\begin{figure} 
  \epsscale{1.15}
  \plotone{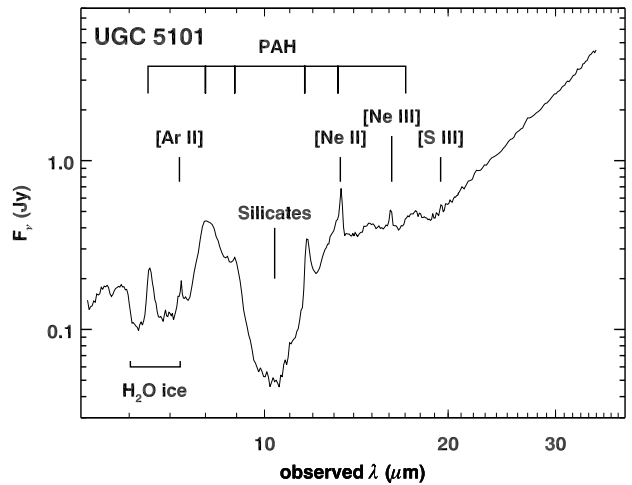}
  \caption{A low-resolution spectrum of UGC\,5101 (using both Short-Low
    and Long-Low), showing several strong bands from PAHs, dust, and ice
    along with atomic emission lines (redshifted for z=0.039).  The
    integration time was 12 seconds in each Short-Low sub-slit and 28
    seconds in each Long-Low sub-slit (total 80 seconds).
    \citet{Armus04} discuss this spectrum more completely.}
\end{figure}

\section{Conclusion} 
The IRS is a major step forward in speed and sensitivity.  It clearly
enables the extension of infrared spectroscopy to a very large number
of extragalactic sources.  The designed spectral resolution is well
matched to the expected line widths from these objects.  The IRS also
allows mid-infrared spectroscopy of objects such as brown dwarfs,
individual stars in neighboring galaxies, and a wide variety of other
sources previously difficult or impossible to study spectroscopically
in the mid-infrared.  This description of the IRS gives the potential
observer a brief review of its design, its capabilities, and how to
use it.  Now it is up to the astronomical community to fully exploit
what the IRS can do.

\acknowledgements 

The design, fabrication, and testing of the IRS at Ball Aerospace
Corporation was supported by a NASA contract awarded by the Jet
Propulsion Laboratory to Cornell University (JPL contract number
960803).  The IRS pipeline was developed at the SSC at the California
Institute of Technology under contract to JPL.

\end{document}